\documentclass[epjCONF]{svjour}

\usepackage{latexsym}
\usepackage{graphics}
\usepackage{graphicx}
\usepackage{epsf,epsfig,amsmath}
\usepackage{bm} 

\usepackage[varg]{txfonts} 
\usepackage[latin1]{inputenc}
\session-title{%
19$^{\textnormal{\footnotesize th}}$ International %
IUPAP Conference on Few-Body Problems in Physics%
}

\begin{document}
\title{Renormalization and Universality of Van der Waals
  forces\thanks{Presented by E. Ruiz Arriola at 19th International
    IUPAP Conference On Few-Body Problems In Physics (FB 19) 31 Aug -
    5 Sep 2009, Bonn, Germany} }
\author{\underline{Enrique Ruiz Arriola}
\thanks{\email{earriola@ugr.es}}
 \and Alvaro Calle Cord\'on\thanks{\email{alvarocalle@ugr.es}} }
%
%
\institute{
Departamento
  de F\'isica At\'omica, Molecular y Nuclear, Universidad de Granada,
  E-18071 Granada, Spain}
\date{Received: date / Revised version: date}
%
\abstract{Renormalization ideas can profitably be exploited in
  conjunction with the superposition principle of boundary conditions
  in the description of model independent and universal scaling
  features of the singular and long range Van der Waals force between
  neutral atoms. The dominance of the leading power law is highlighted
  both in the scattering as well as in the bound state problem.  The
  role of off-shell two-body unitarity and causality within the
  Effective Field Theory framework on the light of universality and
  scaling at low energies is analyzed.}
\maketitle
\section{Introduction}
\label{intro}

Van der Waals (VdW) forces were first conjectured from the
experimental observation that in an adiabatic expansion a gas of
neutral particles cools down (Joule-Thomson effect).  Since the
inter-particle distance at room temperature is $\sim 30 \AA$ this
suggests that VdW forces are {\it long range} and {\it
  attractive}.~\footnote{A very readable historical account can be
  found in Ref.~\cite{2005cohe.book.....R}.}
Their genuine quantum mechanical origin and form $\sim 1/r^6$ was
uncovered by London~\cite{1930ZPhy...63..245L} as long
range dipole fluctuations between charge-neutral atomic and molecular
systems. They dominate at distances above $5-10 \AA$ and hold
atomic dimers together.  The {\it relativistic} Casimir-Polder forces
$\sim 1/r^7$ include retardation, are a consequence of vacuum
fluctuations~\cite{Casmir:1947hx} and operate at very long distances
$\sim 1000-2000 \AA$, a relevant scale in colloids. The general field
theoretical treatment due to two photon
exchange~\cite{Feinberg:1970zz} yielded the so far missing magnetic
contribution (For a review see e.g. \cite{Feinberg:1989ps}).

Van der Waals forces, besides being long range, diverge if directly
extrapolated to short distance scales but a sensible interpretation
becomes possible~\cite{Case:1950an,Frank:1971xx}. Because of the
interest on ultra-cold atoms in recent
years~\cite{2004cucq.book.....W} fundamental work for neutral atoms
was initiated in
Refs.~\cite{PhysRevA.48.546,Gao:1998zza,PhysRevA.59.1998} (see also
\cite{2009PhRvA..80a2702G}) from the point of view of quantum defect
theory, where a spectacular reduction of parameters takes place. This
is supported by more conventional potential calculations and a pattern
of (VdW) universality and scaling sets in~\cite{Cordon:2009vdw} with
no explicit reference to short distance scales or cut-offs.  Actually,
Effective Field Theories (EFT) explicitly exploit the characteristic
low energy parameter reduction from the start and yield very general
universality patterns which do not resolve the nature of the forces
and therefore enjoy a wide
applicability~\cite{Braaten:2004rn,Braaten:2007nq,Platter:2009gz}.
They are based on pure contact (zero range) interactions and discard
the long distance tail of VdW forces. In the present contribution we
analyze the quantum mechanical problem from the point of view of
renormalization, and address to what extent do these contact
interactions faithfully describe the underlying Van der Waals force.

\section{From Binding to Van der Waals forces}

\begin{figure}[ttt]
\epsfig{figure=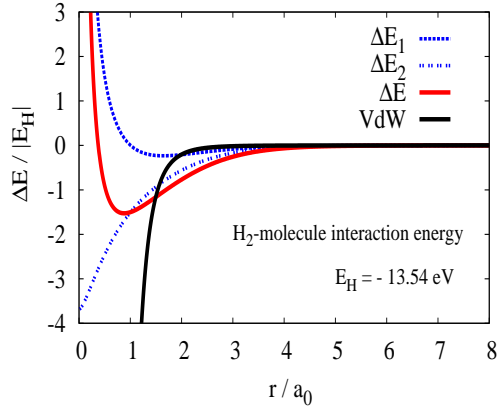,height=6cm,width=7cm,angle=0}
\caption{\small Born-Oppenheimer interaction energies for the $H_2$ molecule
 (in units of the ionization energy of the H atom, $E_H=-m_e
  c^2 \alpha^2/2 = 13.54 {\rm eV}$) as a function of distance (in
  units of the Bohr radius $a_0 = \hbar^2 /m_e c^2 = 0.51 \AA$). We
  compare first order, $\Delta E_1$, and second order, $\Delta E_2$,
  perturbation theory with the pure VdW approximation $\Delta
  E_{\rm VdW}= - 6 m_e c^2 \alpha^2 (a_0/r)^6$. All calculations use
  the closure approximation.}
\label{fig:del1+del2}     
\end{figure}

To provide a proper perspective it is interesting to recall the
distance scales where we expect the dispersion forces to dominate. For
simplicity let us consider the simplest $H_2$ molecule, which
Hamiltonian in the CM frame and in the Born-Oppenheimer approximation,
valid for heavy protons, $m_p \gg m_e $, reads~\footnote{We work in
  natural units with $\hbar=c=1$ and $\alpha = e^2 /(4\pi \epsilon_0
  \hbar c)= \alpha = 1/137.04$ the fine structure constant and $\hbar
  c = 1973.2 \AA {\rm eV }$. The Bohr radius is $a_0 = \hbar^2 /(m_e
  \alpha) = 0.51 \AA$.}
\begin{eqnarray}
H= H_1 + H_2 + V_{12} 
\label{eq:Htotal}
\end{eqnarray}
where the single atom hydrogen-like Hamiltonians are
\begin{eqnarray}
H_{1,2} = -\frac{1}{2 m_e} \vec \nabla_{1,2}^2 -  \frac{\alpha}{|\vec r_{1,2} \pm \vec r/2|}
\label{eq:h-single}
\end{eqnarray}
with $\vec r_1$ and $\vec r_2$ the electron coordinates and 
\begin{eqnarray}
V_{12}=  \alpha 
\left[ \frac{1}{r} + \frac{1}{|\vec r_{1}-\vec r_{2}|} -
 \frac{1}{|\vec r_{1}-\vec r/2|} -
 \frac{1}{|\vec r_{2}+\vec r/2|} \right] 
\end{eqnarray}
Defining $\vec r_{1,\pm} = \vec r_1 \pm \vec r/2$ and $\vec r_{2,\pm}
= \vec r_2 \pm \vec r/2$, the solutions to Eq.~(\ref{eq:h-single}) are
$\psi_n (\vec r_{1,-} )$ and $\psi_m (\vec r_{2,+} )$ where $E_n =
-m_e \alpha^2/2n^2 = -13.6 {\rm eV}/n^2$~\cite{1974-Levine} so that
for $V_{12}=0$ the total mirror symmetric molecular normalized wave
function reads
\begin{eqnarray}
\Psi^{(0)}_{n,m}(\vec r_1, \vec r_2) = \frac{\psi_n (\vec r_{1,-})
  \psi_m (\vec r_{2,+}) \pm \psi_m (\vec r_{1,+}) \psi_n (\vec
  r_{2,-})}{\sqrt{2 (1\pm S_{nm})}} \, , \nonumber \\ 
\label{eq:mol-wf}
\end{eqnarray}
with $S_{nm}(r ) $ the corresponding overlap integral, fulfilling
$S_{n,m} (0)= \delta_{n,m}$ and $S_{n,m}= {\cal O} ( e^{-2 r /a_0 })$.
This generates a coupled channel matrix Hamiltonian what eigenvalues
provide the H-H adiabatic energy, $E_{\rm HH}(r) \to 2 E_{\rm H}$ for
$r \to \infty$.  The nowadays standard variational approach pioneered
by Heitler and London~\cite{1927ZPhy...44..455H} and culminating with
the benchmark determination of the ground state dissociation
energy~\cite{1960RvMP...32..219K} does not accurately work at very
long distances, and thus perturbation theory might be preferable.
Taking $H_0= H_1+H_2$ as the unperturbed Hamiltonian and $V_{12}$ as
the perturbation, one can determine the potential energy shift of the
system at a fixed proton-proton separation $r$ in perturbation theory,
which to second order reads,
\begin{eqnarray}
V_{\rm H-H}(r) \equiv E_{\rm H-H}(r)-2 E_{\rm H}= 
  \Delta E_1(r) + \Delta E_2 (r) + \dots 
\label{eq:del1+del2}
\end{eqnarray}
In the case of the $H_2$ molecule the calculation was undertaken in
1930 by London and Eisenschitz~\cite{1930ZPhy...60..491E} in the
closure approximation (CA)~\footnote{This somewhat crude approximation
  corresponds to replace
$$ \sum_{n\neq 0} \frac{|V_{n,0}|^2}{E_n-E_0} \approx 
\frac{(V^2)_{0,0} -V_{0,0}^2}{E_1-E_0} 
$$ 
Note that the sum includes {\it also} continuum $e-p$ states.}. We have
reproduced the analytical calculation~\cite{Alvaro-Tesina} and the
results are presented in Fig.~\ref{fig:del1+del2}. Already in $\Delta
E_1 (r)$ the finite atomic size yields effects which are $\sim
e^{- 2 r/a_0}$.


An interesting feature is that due to the finite atomic size, $\sim
a_0$, second order in perturbation theory is finite at zero
separation, $|\Delta E_2(0) | < \infty$. Actually, for $r \to 0$ we
expect the exact behaviour $V_{\rm H-H} (r) - \alpha/r \to E_{\rm He}
-2 E_{\rm H}$. In the CA we get $E_{\rm He}^{\rm CA} = -110
{\rm eV} $ to be compared with $E_{\rm He}^{\rm exact} = -79 {\rm eV} $. At
large distances {\it only} the second order direct term in
Eq.~(\ref{eq:del1+del2}) contributes yielding,
\begin{eqnarray}
V (r) = - \sum_{n=6}^\infty  \frac{C_n}{r^n} + {\cal O} (e^{-2 r/a_0}) 
\label{eq:Cn}
\end{eqnarray}
where in the CA $C_n= - c_n m_e c^2 \alpha^2 a_0^n$ with $c_6=6$,
$c_8=135$, $c_{10}=7875/2$, about $5\%$ accurate compared to exact
values (see table~\ref{tab:table_g}). For $n \gg 6$ one has $ C_n \sim
(a_0)^n n!$ which means that the VdW series represents a diverging
asymptotic expansion. As it usually happens in such a case, for a
fixed order of the truncation, this poses a lower limit on the
distance below which it makes no sense improve the
calculation. Usually, only the terms with $n=6,8,10$ are
retained.~\footnote{See e.g. the impressive calculation in Hydrogen up
  to $C_{32}$~\cite{2005PhRvA..71c2709M} for which we can fit a
  behaviour $C_n \sim (1/2)^n n!$ in atomic units. In the CA we have
  $C_{32}^{\rm CA} = 1.93 \times 10^{26}$ to be compared with
  $C_{32}^{\rm exact} = 2.51 \times
  10^{26}$~\cite{2005PhRvA..71c2709M}.}  Calculations show that, so
far, corrections are always negative, i.e. $C_n > 0$ as could be
inferred from the Lieb-Thirring universal
bound~\cite{1986PhRvA..34...40L} which establishes that generally
$V(r) < -c/r^6$ for any pair of atoms with underlying Coulomb forces
such as in Eq.~(\ref{eq:h-single}).  The positivity of the $C_n$'s
will prove essential in what follows. As we see in
Fig.~\ref{fig:del1+del2} in the $H_2$ case the Van der Waals force
dominates for distances of about $5 \AA$. Actually this corresponds to
an energy which is comparable to the environmental translational
thermal energy $k_B T \sim 1/40 \, {\rm
  eV}$~\cite{2003q.bio....12005C}.  Clearly in the ultra-cold region
the VdW force dominates. Therefore, a VdW theory should work in a very
broad range of distances and wavelengths.

The VdW potentials are valid assuming large distances $ r \gg a_0$ and
low energies $p^2 /(2\mu) \ll \Delta = E_1-E_0$ so that intermediate
excited states do not contribute dynamically. 
Thinking of the molecular wave function like e.g.  Eq~(\ref{eq:mol-wf}), at 
large distances we may neglect the exchange term with accuracy 
${\cal O} (e^{-r/a_0})$, and the coupled channel Hamiltonian reads 
\begin{eqnarray} 
H_{n',m'; n,m} (r) = (E_{n}+E_m) \delta_{n',n}
\delta_{m',m} + V_{n',m'; n,m} (r) 
\end{eqnarray}
This problem may quite generally be decomposed within the total
multichannel Hilbert space into the $P$-space (elastic) and $Q$-space
(excited) states~\cite{Lane:1948zh}, where $P=|0,0\rangle \langle
0,0|$ and $Q=1-P$ are the corresponding projection operators. This yields 
the box-matrix structure 
\begin{eqnarray}
H_{PP} \Psi_P + V_{PQ} \Psi_Q  &=& E \Psi_P \nonumber \\ 
V_{QP} \Psi_P + H_{QQ} \Psi_Q  &=& E \Psi_Q 
\end{eqnarray}
where in the Born-Oppenheimer approximation all these potentials are
{\it local} functions of the inter-nuclear separation $r$. Eliminating
the unobserved channels we get the effective optical potential 
\begin{eqnarray}
\bar V_{PP} (E) = V_{PP}  + V_{PQ} (E-H_{QQ})^{-1} V_{QP} 
\end{eqnarray}
which develops an imaginary part if the first inelastic threshold
becomes open. The important point is that if the complete underlying
electronic Hamiltonian, Eq.~(\ref{eq:Htotal}) is energy independent,
then for $E=p^2 /(2\mu)+2 E_{\rm H} < E_{\rm H^*}+ E_{\rm H}$
\begin{eqnarray}
\bar V_{PP}' (E) &=& -V_{PQ} (E-H_{QQ})^{-2} V_{QP} < 0 
\label{eq:positivity}
\end{eqnarray}
The underlying local and energy independent dynamics
has consequences in the low energy EFT representations of the VdW
interactions (see Section \ref{sec:photon-less}).

\section{Universal Scaling Theorems}
\label{sec:universal}

We review here some results found in a previous
work~\cite{PavonValderrama:2005wv,PavonValderrama:2007nu} (for a short
review see e.g. \cite{RuizArriola:2007wm}),within a nuclear physics
and multichannel context which will prove useful in the analysis of
VdW forces
. Our starting point is the finite energy scattering
state Schr\"odinger's equation for the relative wave function between
two particles of masses $m_1$ and $m_2$ which interact through a
central potential,
 \begin{eqnarray}
-u'' + U(r) u + \frac{l(l+1)}{r^2}u= k^2 u \, , 
\label{eq:schr} 
\end{eqnarray} 
where $U(r) = 2 \mu V(r) $, $\mu = m_1 m_2 /(m_1 +m_2) $ is the
reduced mass, $k=p/\hbar =2\pi/ \lambda $ the wavenumber and $u(r)$
the reduced wave function. We will neglect finite size and exchange
effects and take $V(r)$ given just by Eq.~(\ref{eq:Cn}) where $C_n$
are the standard VdW coefficients which are computed {\it ab initio}
from electronic orbital atomic structure
calculations~\cite{KMS-data}. The potential in Eq.~(\ref{eq:Cn}) can
conveniently be rewritten as
\begin{eqnarray}
U(r) = - \frac{R_6^4}{r^6} \left[ 1 + g_1\frac{R_6^2}{r^2}+ 
g_2 \frac{R_6^4}{r^4} +  \dots  \right] \, , 
\label{eq:C6-C8-C10}
\end{eqnarray} 
where $ R_6 = ( 2 \mu C_6 )^\frac{1}{4}$ is the VdW length scale and
$g_1$, $g_2$, etc. represent the contributions from $C_8$, $C_{10}$
etc. at $r=R_6$ respectively. In table~\ref{tab:table_g} we compile
numbers for a bunch of interesting cases. Typically, $R_6 \sim 10-200
\AA$ but also $g_1 \sim 10^{-2}$ and $g_2 \sim 10^{-4}$. This raises
immediately the question under what conditions the expansion
(\ref{eq:C6-C8-C10}) can be truncated in a meaningful way, i.e. when
the neglected terms can indeed be considered negligible in scattering
and bound state properties. This question is intriguing since at short
distances the more singular terms are manifestly more
divergent. Actually, the range where $C_8$ yields an important
correction but $C_{10}$ is still small is $ g_2^{-1/4} \ll r/R_6 \ll
g_1^{-1/2}$ which in view of table~\ref{tab:table_g}, $g_2 \sim
g_1^2$, becomes extremely narrow or inexistent.

\begin{table*}[ttt]
\caption{\label{tab:table_g} Reduced di-atomic masses, Van der Waals
  dispersion coefficients $C_6$, $C_8$ and $C_{10}$ as well as the
  leading Van der Waals length scale $R_6 =( 2 \mu C_6 )^\frac14 $,
  and the dimensionless coefficients $g_1$ and $g_2$ defined by the
  dimensionless reduced potential $2 \mu V(r) R_6^2 \equiv R_6^2 U(r)
  = - x^{-6} \left[ 1 + g_1 x^{-2}+ g_2 x^{-4} + \dots \right] $ with
  $x=r/R_6$ the distance in Van der Waals units, of the di-atomic
  systems used in the present paper. Atomic units are used
  throughout. The atomic masses have been taken from the National
  Institute of Standards and Technology
  http://physics.nist.gov/PhysRefData/Compositions/}
\begin{center}
\begin{tabular}{lccccccc} 
\hline\noalign{\smallskip}
{\rm Atoms} &$\mu(a.u.)$ &$C_6 (a.u.)$&$C_8 (10^{5} a.u.)$&$C_{10} (10^{7}a.u.)$&$R_6(a.u.)$&$g_1 (10^{-2})$&$g_2 (10^{-4})$ \\ 
\noalign{\smallskip}\hline\noalign{\smallskip} 
H-H       &    918.576 &     6.499~\cite{1996PhRvA..54.2824Y} & 0.001244~\cite{1996PhRvA..54.2824Y} &  0.0003286~\cite{1996PhRvA..54.2824Y} &  10.4532    &   17.51760    &  423.45441    \\ 
He-He     &   3648.150 &     1.461~\cite{1996PhRvA..54.2824Y} & 0.000141~\cite{1996PhRvA..54.2824Y} &  0.00001837~\cite{1996PhRvA..54.2824Y} &    10.1610    &    9.35937    &  117.94642 \\ 
Li-Li     &  6394.697  &  1389.~\cite{2001PhRvA..63e2704D}  &   0.834~\cite{2003JChPh.119..844P}  &  0.735~\cite{2003JChPh.119..844P}   &    64.9214    &    1.42458    &    2.97874    \\ 
Na-Na     &  20953.894 &  1556.~\cite{2001PhRvA..63e2704D}  &   1.160~\cite{2003JChPh.119..844P}  &  1.130~\cite{2003JChPh.119..844P}   &    89.8620    &    0.92320    &    1.11369    \\ 
K-K       &  35513.247 &  3897.~\cite{2001PhRvA..63e2704D}  &   4.200~\cite{2003JChPh.119..844P}  &  5.370~\cite{2003JChPh.119..844P}   &   128.9846    &    0.64780    &    0.49784    \\ 
Rb-Rb     &  77392.363 &  4691.~\cite{2001PhRvA..63e2704D}  &   5.770~\cite{2003JChPh.119..844P}  &  7.960~\cite{2003JChPh.119..844P}   &   164.1528    &    0.45647    &    0.23370    \\ 
Cs-Cs     & 121135.907 &  6851.~\cite{2001PhRvA..63e2704D}  &  10.200~\cite{2003JChPh.119..844P}  & 15.900~\cite{2003JChPh.119..844P}   &   201.8432    &    0.36544    &    0.13983    \\ 
Fr-Fr     & 203270.053 &  5256.~\cite{2001PhRvA..63e2704D}  &   6.648~\cite{1998PhRvA..58.4259M}  & 10.699~\cite{1998PhRvA..58.4259M}   &   215.0006    &    0.27362    &    0.09526    \\ 
Li-Na     &   9798.954 &  1467.~\cite{2001PhRvA..63e2704D}  &   0.988~\cite{2003JChPh.119..844P}  &  0.916~\cite{2003JChPh.119..844P}   &    73.2251    &    1.25605    &    2.17183    \\ 
Li-K      &  10837.871 &  2322.~\cite{2001PhRvA..63e2704D}  &   1.950~\cite{2003JChPh.119..844P}  &  2.100~\cite{2003JChPh.119..844P}   &    84.2285    &    1.18374    &    1.79689    \\ 
Li-Rb     &  11813.296 &  2545.~\cite{2001PhRvA..63e2704D}  &   2.340~\cite{2003JChPh.119..844P}  &  2.610~\cite{2003JChPh.119..844P}   &    88.0587    &    1.18572    &    1.70555    \\ 
Li-Cs     &  12148.102 &  3065.~\cite{2001PhRvA..63e2704D}  &   3.210~\cite{2003JChPh.119..844P}  &  3.840~\cite{2003JChPh.119..844P}   &    92.8950    &    1.21364    &    1.68241    \\ 
Na-K      &  26356.596 &  2447.~\cite{2001PhRvA..63e2704D}  &   2.240~\cite{2003JChPh.119..844P}  &  2.530~\cite{2003JChPh.119..844P}   &   106.5708    &    0.80600    &    0.80155    \\ 
Na-Rb     &  32978.812 &  2683.~\cite{2001PhRvA..63e2704D}  &   2.660~\cite{2003JChPh.119..844P}  &  3.130~\cite{2003JChPh.119..844P}   &   115.3377    &    0.74528    &    0.65923    \\ 
Na-Cs     &  35727.672 &  3227.~\cite{2001PhRvA..63e2704D}  &   3.620~\cite{2003JChPh.119..844P}  &  4.550~\cite{2003JChPh.119..844P}   &   123.2277    &    0.73874    &    0.61148    \\ 
K-Rb      &  48685.873 &  4274.~\cite{2001PhRvA..63e2704D}  &   4.930~\cite{2003JChPh.119..844P}  &  6.600~\cite{2003JChPh.119..844P}   &   142.8292    &    0.56543    &    0.37106    \\ 
K-Cs      &  54924.387 &  5159.~\cite{2001PhRvA..63e2704D}  &   6.620~\cite{2003JChPh.119..844P}  &  9.400~\cite{2003JChPh.119..844P}   &   154.2909    &    0.53903    &    0.32152    \\ 
Rb-Cs     &  94444.928 &  5663.~\cite{2001PhRvA..63e2704D}  &   7.690~\cite{2003JChPh.119..844P}  & 11.300~\cite{2003JChPh.119..844P}   &   180.8480    &    0.41520    &    0.18654    \\ 
Cr-Cr     &  47340.881 &   733.~\cite{2005PhRvL..94r3201W}  &   0.750~\cite{2005PhRvL..94r3201W}  &    $-$    &    91.2731    &    1.22821    &    $-$        \\\noalign{\smallskip}\hline
\end{tabular}
\end{center}
\end{table*}
To determine the solution of Eq.~(\ref{eq:schr}) it is necessary to
give sensible boundary conditions at the origin and infinity. For the
usual regular potentials there are a regular and irregular solution at
the origin, and the regularity condition $u(0)=0$ fixes uniquely the
solution. However, since the potential is singular and attractive
there are two linearly independent solutions, so regularity at
the origin does not select a unique solution. Indeed, at short
distances the De Broglie wavelength is slowly varying, $d
[U(r)]^{-\frac12} /dr \ll 1$ and hence a WKB approximation
holds~\cite{Case:1950an,Frank:1971xx}, yielding for $ r \to 0 $
\begin{eqnarray}
u_k (r) &\to & C \left(\frac{r}{R_n}\right)^{n/4} \sin\left[
\frac{2}{n-2} \left(\frac{R_n}{r}\right)^{\frac{n}2-1} + \varphi_k \right] \, ,  \label{eq:WKB} 
\end{eqnarray} 
where $R_n= (2\mu C_n)^{1/(n-2)}$ corresponds to the highest VdW scale
considered in Eq.~(\ref{eq:Cn}) (see also
Eq.~(\ref{eq:C6-C8-C10}). The phase $\varphi_k $ is arbitrary and
could, in principle, be energy dependent (see below.

To fix ideas we will restrict to $l=0$, $s-$waves. 
For a positive energy scattering state it is convenient to use the
normalization at very long distances given by
\begin{eqnarray}
u_k (r) \to \frac{\sin(kr + \delta_0)}{\sin \delta_0} = \cos(kr) + k
\cot \delta_0 \frac{\sin (kr)}{k} \, , 
\label{eq:long} 
\end{eqnarray} 
where $\delta_0(k) $ is the scattering phase shift for the $l=0$
angular momentum state. For the potential which at long distances
behave as Eq.~(\ref{eq:C6-C8-C10}) one has the effective range
expansion (ERE)~\cite{1963JMP.....4...54L}
\begin{eqnarray}
k \cot \delta_0 (k)= - \frac1{\alpha_0} + \frac12 r_0 k^2 + v_2 k^4
\log (k^2) + \dots
\label{eq:ere} 
\end{eqnarray} 
where $\alpha_0$ is the scattering length, and $r_0$ is the effective
range. $\alpha_0$ and $r_0$ can be calculated from
the asymptotic behaviour of the zero energy solution
\begin{eqnarray}
u_0 (r) \to 1 - r/\alpha_0  \, , 
\label{eq:a0_singlet}
\end{eqnarray} 
and using the definition
\begin{eqnarray} 
r_0 &=& 2 \int_0^\infty dr \left[ \left(1-r/\alpha_0 \right)^2-
u_0 (r)^2 \right] \, . 
\label{eq:r0_singlet}
\end{eqnarray} 
Next, we use the superposition principle of boundary conditions and
write
\begin{eqnarray}
u_k (r) = u_{k,c} (r)  +  k \cot \delta_0 \,  u_{k,s} (r)    \, , 
\label{eq:sup_k}
\end{eqnarray} 
with $ u_{k,c} (r) \to \cos (k r) $ and $ u_{k,s} (r) \to \sin (k r)
/k $ for $r \to \infty $. At zero energy we have
\begin{eqnarray}
u_0 (r) = u_{o,c} (r)  - u_{0,s} (r) / \alpha_0    \, , 
\label{eq:sup_0}
\end{eqnarray} 
with $ u_{0,c} (r) \to 1 $ and $ u_{0,s} (r) \to \sin r $ for $r \to
\infty $.  The short distance phase $\varphi_0$ can be fixed in
practice introducing a {\it short distance} cut-off, $r_c$. The way to
proceed is as follows. Given a scattering length $\alpha_0$ as {\it
  input} one integrates in from large down to small distances, say
$r=r_c \ll R_6 $ whence determining $\varphi_0$. To determine
$\varphi_k$ we do the same but for finite energy states. A relation
between both short distance phases can be found as
follows~\cite{PavonValderrama:2005wv}. If  we build $ (u_k' u_0 - u_0' u_0)'$ 
and integrate from $r_c$ to infinity and  use 
Eq.~(\ref{eq:sup_k}) and Eq.~(\ref{eq:sup_0}) respectively, one gets 
\begin{eqnarray}
\frac1{R_n} \sin( \varphi_k -\varphi_0) &=& k^2 \int_0^\infty dr 
\left[u_{0,c} (r) - \frac{1}{\alpha_0} \, u_{0,s} (r) \right]
\nonumber \\ & & \times \Big[ u_{k,c} (r) + k \cot \delta_0 (k) \,
  u_{k,s} (r) \Big] \, , 
\label{eq:ortho}
\end{eqnarray} 
which becomes an orthogonality relation if and only if
$\varphi_k=\varphi_0$. This implies that the corresponding
Hamiltonian, although unbounded from below becomes self-adjoint on the
domain of square integrable functions which have the short distance
behaviour, Eq.~(\ref{eq:WKB}), with the {\it same } short distance
phase (a common domain of definition). In other words, the short
distance common phase, $\varphi$, labels the particular self-adjoint
extension of the Hamiltonian depending on the Van der Waals dispersion
coefficients.~\footnote{The energy independence can also be deduced
  from the smallness of the wave function at small distances.}.  One
important property is that $\varphi$ is not only energy independent
but it must be fixed independently on the long distance potential, and
thus encodes all short distance information not accounted for by the
truncated potential Eq.~(\ref{eq:C6-C8-C10}). Moreover, there is a
further remarkable consequence of the energy independence of
$\varphi$ obtained by  expanding the integrand in
Eq.~(\ref{eq:ortho}), 
\begin{eqnarray}
k \cot \delta_0 (k) = \frac{ \alpha_0 {\cal A} ( k) + {\cal B} (k)}{
\alpha_0 {\cal C} (k) + {\cal D} (k)} \, , 
\label{eq:phase_singlet}
\end{eqnarray} 
whereas the functions ${\cal A}$, ${\cal B}$, ${\cal C}$ and ${\cal
D}$ are even functions of $k$ which depend {\it only on the potential}
and are given by
\begin{eqnarray}
{\cal A}(k) &=& \int_0^\infty  dr \, u_{0,c} (r)  u_{k,c} (r) \, ,\\    
{\cal B}(k) &=& \int_0^\infty  dr \, u_{0,s} (r)  u_{k,c} (r) \, ,\\
{\cal C}(k) &=& \int_0^\infty  dr \, u_{0,c} (r)  u_{k,c} (r) \, ,\\ 
{\cal D}(k) &=& \int_0^\infty  dr \, u_{0,s} (r)  u_{k,s} (r) \, . 
\end{eqnarray} 
Note that the dependence of the phase-shift on the scattering length
is wholly {\it explicit}; $\cot \delta_0 $ is a bilinear rational
mapping of $\alpha_0$. This is just a manifestation of the underlying
Moebius transformation well known from the theory of ordinary
differential equations discussed in~\cite{PavonValderrama:2007nu}. The
obvious conditions ${\cal A}(0)={\cal D}(0)=0$ and ${\cal B}(0)={\cal
  C}(0)=1$ are satisfied. The limiting procedure poses no problem in
principle, and the divergence of the potential at short distances has
been eliminated by {\it demanding} a finite physical scattering
length. This is equivalent to a renormalization condition at zero
energy. We will analyze below other alternative renormalization
conditions based on one bound state.  The effective range is defined
by Eq.~(\ref{eq:r0_singlet}), and using Eq.~(\ref{eq:sup_0}) we get
\begin{eqnarray} 
r_0  &=&  A + \frac{B}{\alpha_0}+ \frac{C}{\alpha_0^2}  \, ,    
\label{eq:r0_univ} 
\end{eqnarray} 
where 
\begin{eqnarray}
A &=& 2 \int_0^\infty dr ( 1 - u_{0,c}^2 ) \, , \\    
B &=& -4 \int_0^\infty dr ( r - u_{0,c} u_{0,s} ) \, , \\    
C &=& 2 \int_0^\infty dr ( r^2 - u_{0,s}^2 )    \, ,  
\end{eqnarray} 
depend on the potential parameters only. The interesting feature of
the previous equations is that all explicit dependence on the
scattering length $\alpha_0$ is displayed by Eq.~(\ref{eq:r0_univ}
). This is a universal form of a low energy theorem, which applies to
{\it any} potential regular or singular at the origin which falls off
faster than $1/r^5$ at large distances. Since the potential is known
accurately at long distances we can visualize Eq.~(\ref{eq:r0_univ})
as a long distance (VdW) correlation between $r_0$ and
$\alpha_0$.  

The above results for the effective range and phase-shift,
Eq.~(\ref{eq:r0_univ}) and Eq.~(\ref{eq:phase_singlet}) are completely
general. If, in addition, the reduced potential depends on a {\it
single } scale $R$, i.e. $U(r) = - F( r/R) /R^2 $, one gets {\it
universal scaling relations}
\begin{eqnarray} 
\frac{r_0}{R} = \bar A + \bar B \frac{R}{\alpha_0}+ \bar C
\frac{R^2}{\alpha_0^2} \, ,
\label{eq:r0_univ_scaled} 
\end{eqnarray} 
and 
\begin{eqnarray} 
R k \cot \delta_0 ( k R , \alpha_0 /R) = \frac{ \alpha_0 \, {\cal A} ( k
R ) + R \, {\cal B} (k R)}{ \alpha_0 \, {\cal C} ( k R) + R \, {\cal D} (k R)} \, , 
\label{eq:phase_scaled}
\end{eqnarray} 
where $\bar A$,$\bar B$ and $\bar C$ are purely geometric numbers, and
${\cal A}$ ,${\cal B}$, ${\cal C}$ and ${\cal D}$ are functions which
depend {\it solely} on the functional form of the potential. Thus, if
a potential has a single scale the phase-shift can be computed {\it
  once and forever} for a given energy. This allows to compare quite
different physical systems which have identical long range forces but
different short distance dynamics. A remarkable consequence of
Eq.~(\ref{eq:r0_univ_scaled}) is that if $\alpha_0 \gg R $ then $r_0
\sim R$ whereas $\alpha_0 \ll R $ implies $r_0 \gg R$.

\section{The Pure $-1/r^6$ VdW force}

\begin{figure}[ttt] 
\centering 
\includegraphics[width=5.5cm,angle=270]{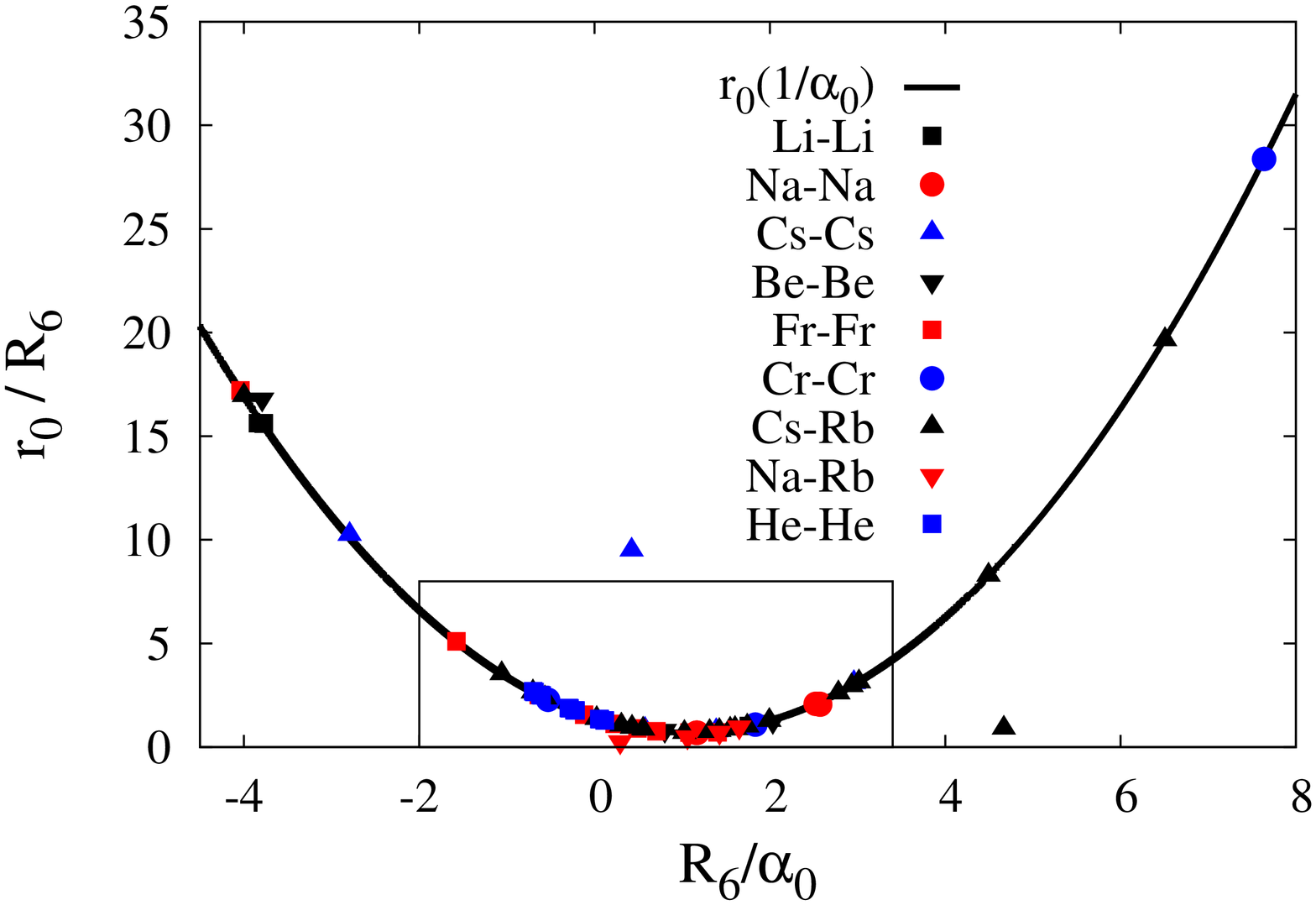} \\
\includegraphics[width=5.2cm,angle=270]{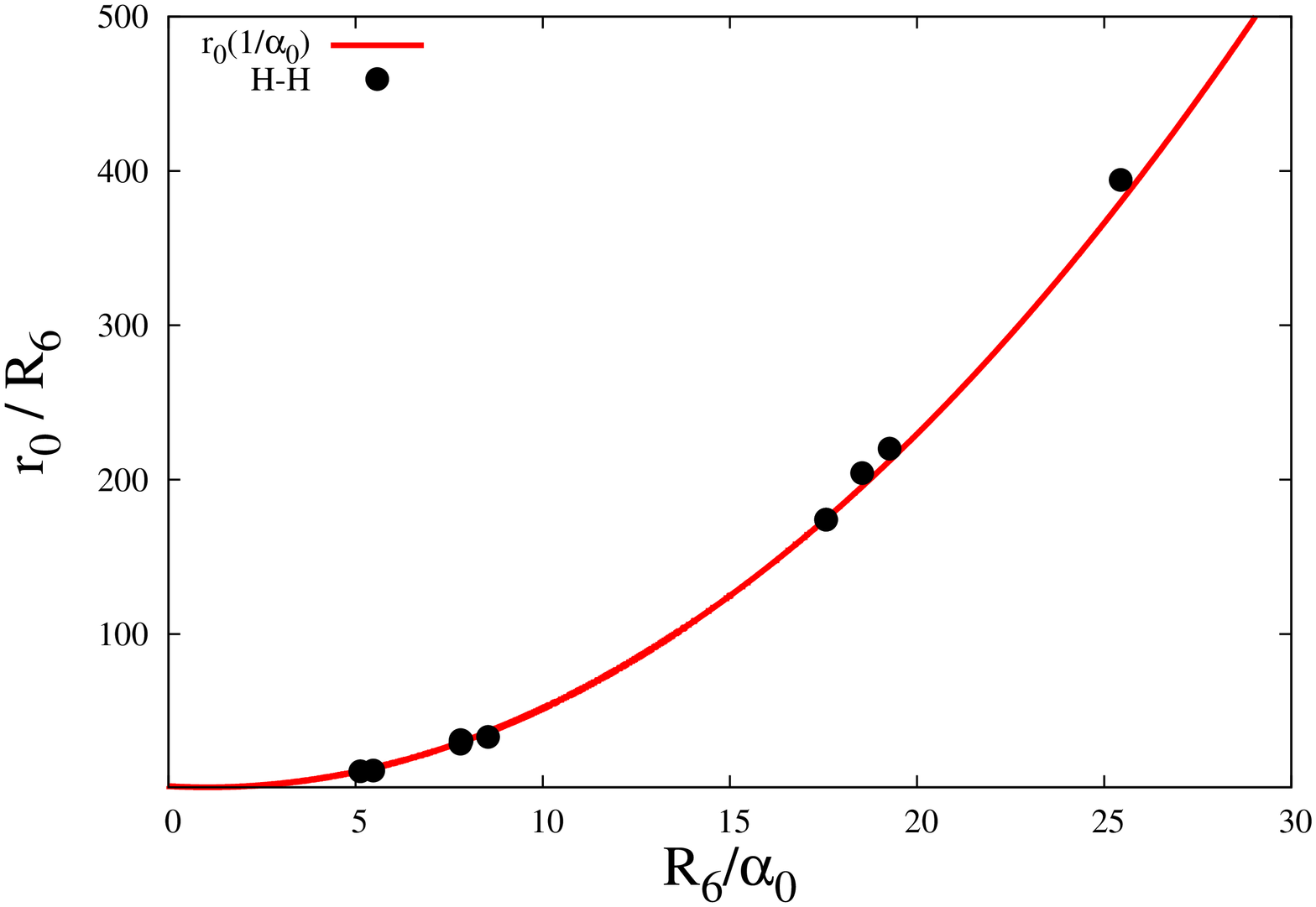} 
\caption{\small The VdW universal effective range in units of the VdW
  radius $R$ defined by $ 2\mu V(r) = -R^4/r^6$. Compared to several
  calculations~\cite{1996PhRvA..53..234C} (Li-Li,Na-Na),
\cite{1994PhRvA..50.4827C} (Na-Na), \cite{1999PhRvA..59.1998F} (Cs-Cs),
\cite{1994PhRvA..50.3177M} (Cs-Cs), \cite{2009EPJD...53...27O} (Na-Rb),
\cite{2007JPhB...40.3497J} (Be-Be), \cite{2003JPhB...36.1085J} (Cs-Rb),
\cite{2004PhRvA..69c0701P} (Cr-Cr), \cite{2004physics...6027K} (Fr-Fr),
\cite{1986JPSJ...55..801K} (H-H), \cite{1992PhRvA..46.6956J} (H-H),
\cite{2006EL.....76..582S} (H-H),  \cite{1995PhRvA..51.2626J} (H-H).} 
\label{fig:atoms-curve}
\end{figure}

\begin{figure}
\begin{center}
\epsfig{figure=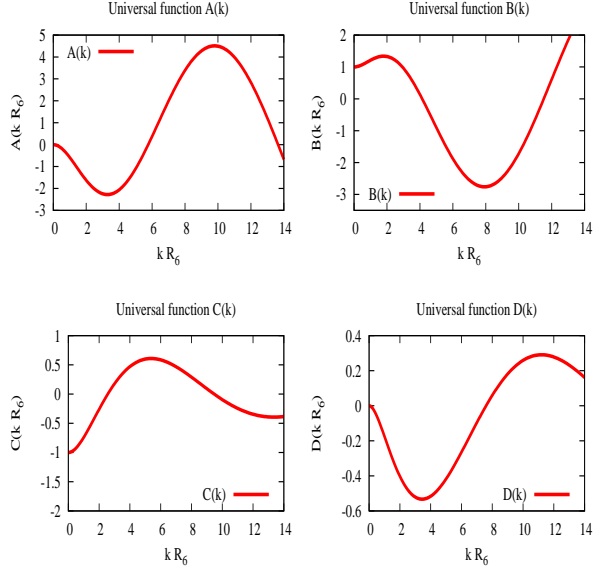,height=8cm,width=8cm,angle=270}
\end{center}
\caption{\small The VdW universal functions in units of the
  VdW radius $R$ defined by $ 2\mu V(r) = -R^4/r^6$. Using
  these functions one can determine the phase shift if the scattering
  length is known by the formula $ k \cot \delta_0 (k) = [ \alpha_0
    {\cal A} ( k R ) + {\cal B} (k R )]/[ \alpha_0 {\cal C} ( k R) +
    {\cal D} (k R)] $.}
\label{fig:ABCD}       
\end{figure}

For the pure $1/r^6$ case one has an analytical solution for the zero
energy state so that the effective range can be computed
analytically~\cite{Gao:1998zza,PhysRevA.59.1998} using
Eq.~(\ref{eq:r0_singlet}) to get
\begin{eqnarray}
\bar A= \frac{16\,\Gamma\left(\frac54\right)^2}{3\pi} \, , \quad  \bar B=- \frac{4}{3} \, , 
\quad \bar C= \frac{4\Gamma\left( \frac34 \right)^2}{3\pi} \, , 
\label{eq:r0-ggf}
\end{eqnarray}
in agreement with the general result, Eq.~(\ref{eq:r0_univ_scaled}).
In Fig.~\ref{fig:atoms-curve} we confront the prediction for the
effective range to the result of many {\it ab initio} calculations.
As we see the agreement is rather impressive, meaning that for many
practical purposes Eq.~(\ref{eq:r0-ggf}) and
Eq.~(\ref{eq:r0_univ_scaled}) summarize the relevant information on
the, about a hundred, data. In Fig.~\ref{fig:ABCD} we show the VdW
universal functions ${\cal A}(k)$, ${\cal B}(k)$, ${\cal C}(k)$ and
${\cal D}(k)$ uniquely determined by the VdW potential and which need
that the scattering length be specified separately to obtain the phase
shift (see Eq.~(\ref{eq:phase_scaled}).

\begin{figure}
\centering 
\includegraphics[width=4.5cm,angle=270]{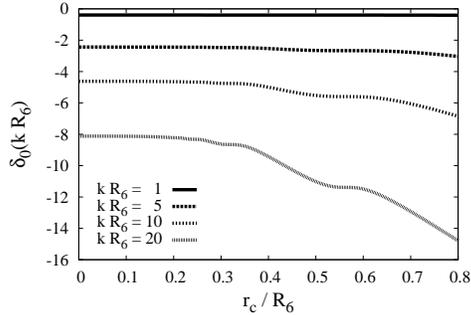}  
\caption{\small Convergence of the phase shift for a fixed momentum
  $k$ as a function of the short distance cut-off, $r_c$ in VdW
  units.}
\label{fig:convergenge}
\end{figure}

One important issue has to do with the cut-off dependence.  From a
Callan-Symanzik renormalization group type argument one
has~\cite{PavonValderrama:2007nu}
\begin{eqnarray}
\frac{d\delta(k,r_c)}{d\,r_c} = -\,k^3 \left( \frac{u_k(r_c,r_c)}{k} \right)^2
\end{eqnarray} 
when $\alpha_0$ is fixed. This in turn means that for $U(r) \sim -
R_6^4/r^{6}$ and using $u_k(r_c,r_c)/k \sim (r_c/R_6)^{3/2}/k$ 
one has that finite cut-off corrections scale as 
\begin{eqnarray}
\delta (k) - \delta(k,r_c)  = {\cal O} (k^3  r_c^4 /R_6 )  
\end{eqnarray}
Thus, the short distance cut-off $r_c$ is a {\it parameter} for $ k
R_6 \gg 1$ but becomes innocuous otherwise when $r_c \ll R_6$ implying
that universality is robust.  We show for a sample case in
Fig.~\ref{fig:convergenge} the phase shifts for fixed energies, where
the rather smooth converging pattern can be clearly confirmed.

\begin{figure}
\centering 
\includegraphics[width=6.5cm,angle=0]{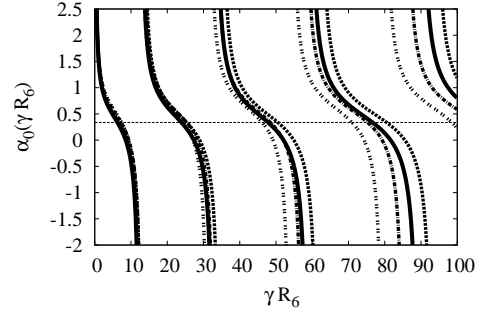}  
\caption{Bound state spectrum for the VdW potential $R_6^2 U(r) = -
  (R_6/r)^6[ 1 + g (R_6/r)^2 ] $ for $g=0$ (solid), $g=0.1$
  (long-dashed-dotted) )$,0.5$ (dashed) and $g=1$
  (short-dashed-dotted). The horizontal line corresponds to a
  situation where $\alpha_0 /R_6=0.336$ and the intersections to
  $\gamma_n$ values with $ E_n = - \gamma_n^2 /(2\mu)$.}
\label{fig:univer-spec}
\end{figure}

Turning to the negative energy bound states with $E=-\hbar^2 \gamma^2
/ (2 \mu)$, the behaviour at long distances is
\begin{eqnarray} 
u_\gamma (r) \to A e^{-\gamma r} 
\end{eqnarray} 
whereas at short distances the energy dependent boundary condition,
Eq.~(\ref{eq:WKB}) , holds.  Orthogonality to the zero energy state
requires that
\begin{eqnarray}
\alpha_0 = \frac{\int_0^\infty dr \, u_\gamma(r) u_{0,s}(r)}
{\int_0^\infty dr \, u_\gamma(r) u_{0,c}(r)} \, .
\end{eqnarray}  
This relation explicitly yields the scattering length from a given
bound state. Inverting the formula we get, the energy
spectrum, $E_n = - \hbar^2 \gamma_n^2 / 2 \mu $ for a fixed value of
the scattering length and a fixed potential~\footnote{Using the WKB
method one obtains (see also Ref.~\cite{PavonValderrama:2007nu}) 
$$
n + c = \left[\frac{1}{2\pi} +\frac{ 3\Gamma \left( \frac53 \right)} {8
\sqrt{\pi} \Gamma \left( \frac76 \right)} \right] ( \gamma_n R
)^{\frac23} = 0.20587 ( \gamma_n R )^{\frac23}
$$
where $c $ is a constant of order unity.  Thus if we happen to have a
zero energy state, $\gamma_0=0$ then $c=0$ and the bound states are
$E_n = - 116.08 n^3 \, \hbar^2 /(2 \mu R^2)$ counting downward in the
spectrum.}.  This relation determines the energy spectrum $E_n $ from
the scattering length $\alpha_0$ and the potential. Conversely, we can
get the scattering length $\alpha_0$ from a given bound state wave
number $\gamma$. In Fig.~\ref{fig:univer-spec} we display the lowest
bound states as functions of $\alpha_0$ for the pure VdW potential in
scaled units. Such an universal plot allows also to deduce the
scattering length from the knowledge of the weakly bound states in a
complete {\it model independent} way.

\begin{figure}[ttt] 
\centering 
\includegraphics[width=6.5cm,angle=0]{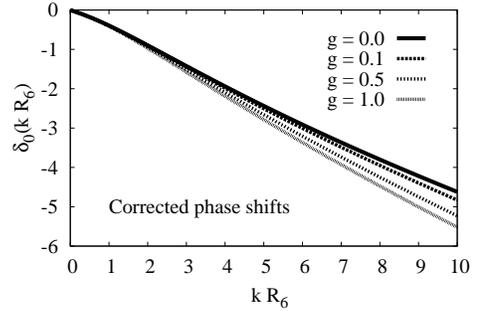} 
\caption{\small s-wave phase shifts for the VdW potential $R_6^2 U(r) = -
  (R_6/r)^6[ 1 + g (R_6/r)^2 ] $ for $g=0,0.1,0.5,1.$ for a fixed
  sample value of the scattering length $\alpha_0 /R_6=0.335$.}
\label{fig:U6+U8}
\end{figure}

We may ask how should the neglected higher order $1/r^8$ ,$1/r^{10}$,
etc.  terms in the potential affect the lowest order $1/r^6$
calculation. To this end we show in Fig.~\ref{fig:U6+U8} the effect of
adding the term $1/r^8$ to the potential, see
Eq.~(\ref{eq:C6-C8-C10}), in the s-wave phase shift fixing the
scattering length $\alpha_0$ to the {\it same} value. In view of the
rather small $g_1 \sim 10^{-2}$ values listed in table
\ref{tab:table_g} we expect tiny corrections at even large energies $k
R_6 \sim 10$. This result not only shows a clear dominance of the
leading dispersion coefficient $C_6$ but also that scaling of VdW
forces holds beyond naive dimensional estimates, $ k R_6 \approx 1$.
This raises the question on the usefulness of including higher order
dispersion coefficients such as $C_8$ and $C_{10}$ since the region
where they can distinctly be disentangled without entering the finite
size regime is extremely narrow. This is also seen in the bound state
spectrum displayed in Fig.~\ref{fig:univer-spec} since for $\gamma R_6
< 10$ a rather tiny energy shift is observed in the closest states to
the continuum.

\section{Phenomenological potentials}
\label{sec:phenomenological}

In our way of treating the renormalization of VdW forces, we need not
specify the value of the scattering length, $\alpha_0$, since it
exactly factors out in the expression for the phase shift (see
Eq.~(\ref{eq:phase_scaled})). However, to predict scattering phase
shifts a particular value of $\alpha_0$ must be used. It is
interesting to analyze the results from a comparative perspective with
the so called realistic inter-atomic potential models, which aim at a
description through the entire range of distances. Thus, one can
deduce from those the value of the scattering length. The advantage of
such potentials is that they provide a complete description of the
interaction throughout the entire separation range. However, many of
its features cannot be deduced accurately from first principles
calculations. In contrast, the long range part of the interaction has
a well accepted form in terms of a few parameters, say $C_6$, $C_8$,
$C_{10}$ which, in principle, can be evaluated from {\it ab initio}
atomic structure electronic wave functions.

Most modern inter-atomic potentials include the asymptotic Van der
Waals long distance behaviour. They are generally written as the sum
of a long range dispersive term and a short distance term with a core
which reflects the impenetrability of two atoms. To keep the
discussion as simple as possible in terms of the number of parameters,
and for illustration purposes, we will analyze the venerable 
Lennard-Jones potential, which we write as 
\begin{eqnarray}
2 \mu V_{\rm L.J.}= U_{LJ} (r) = \frac{1}{R_6^2} \left[ 
g^6
\left(\frac{R_{6}}{r}\right)^{12}-\left(\frac{R_6}{r}\right)^6 \right] \, , 
\end{eqnarray} 
where we have chosen to scale the potential in the long distance VdW
units $R=R_6 = ( 2\mu C_6)^\frac14$. The value of the dimensionless
constant $g$ determines the classical turning point $U_{\rm LJ} ( g
R_6 )=0$. In these form the minimum of the potential is at $r_{\rm
  min}= 2^\frac16 R_6 $ and $U_{\rm min} = - 1/( 4g^6 R_6^2) $. A
relevant dimensionless parameter is the total number of bound states,
which obviously increases as the repulsive term is shifted towards the
origin. Within a WKB approximation the number of bound states is given
by 
\begin{eqnarray}
N_{\rm WKB}= \frac1\pi \int_a^\infty dr \sqrt{-U_{\rm LJ}(r)} =
\frac{0.1339}{g^2} \, , 
\end{eqnarray} 
where $a=g R_6$ is the zero energy classical turning point. For
$g=0.0365$ we get $N_{\rm WKB}=100$, for $g=0.0517491$ we get $N_{\rm
  WKB}=50$ and for $g=0.115715$ we get $N_{\rm WKB}=10$.

\begin{figure}[ttt]
\begin{center}
\epsfig{figure=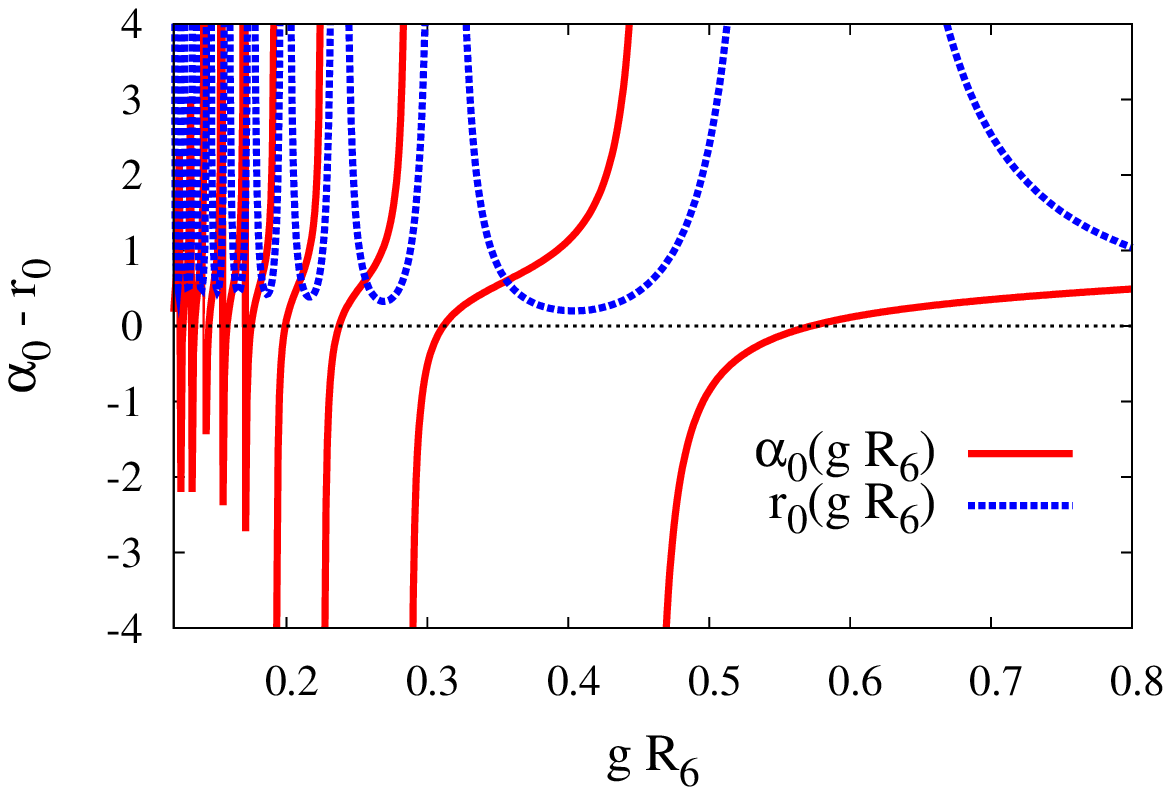,height=4.5cm,width=7cm}
\epsfig{figure=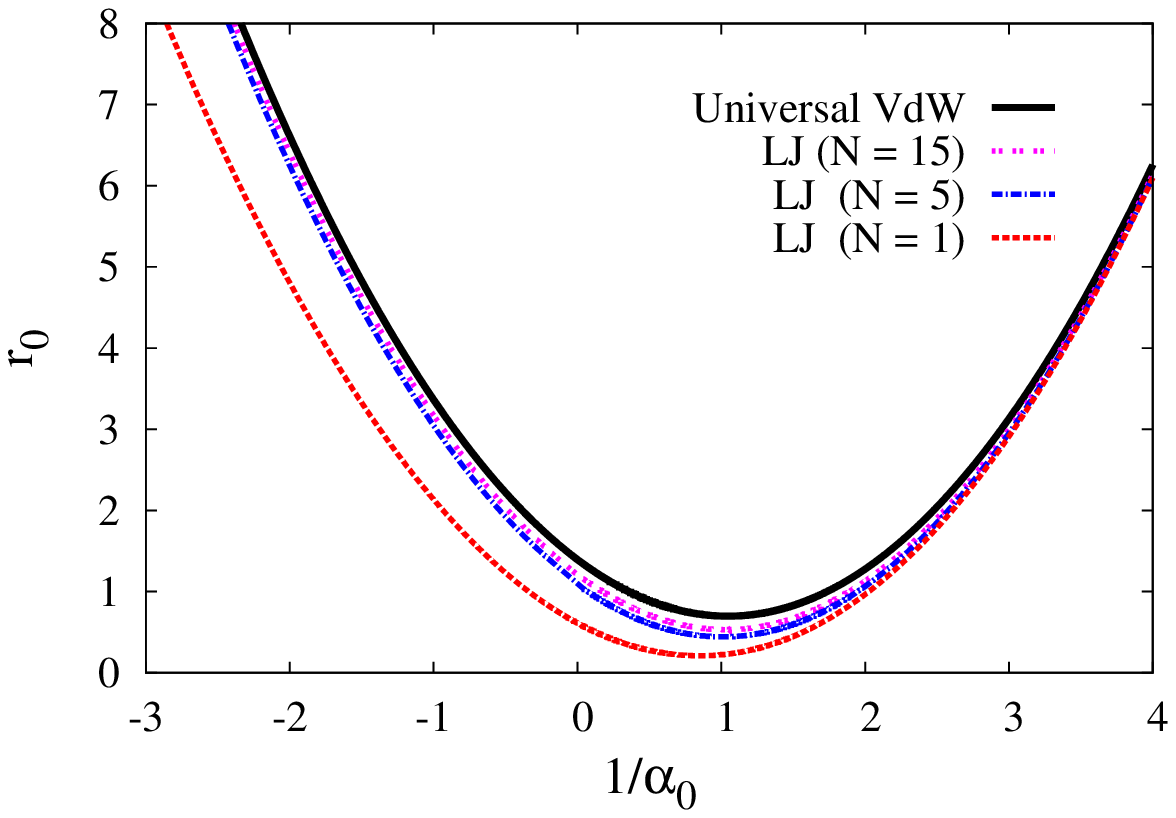,height=4.5cm,width=7cm}
\end{center}
\caption{
(Upper panel) 
The scattering length $\alpha_0$ and effective range $r_0$ of the
Lennard-Jones potential as a function of the zero energy turning point
$g R_6$ (in VdW units).
(Lower panel) The effective range $r_0 $ of the Lennard-Jones model as
a function of the inverse scattering length $1/\alpha_0 $ compared to
the universal renormalized VdW formula (in VdW units) for different
number of bound states  $N=1,5,15$.}
\label{fig:LJ}
\end{figure}

\begin{figure*}[ttt]
\begin{center}
\epsfig{figure=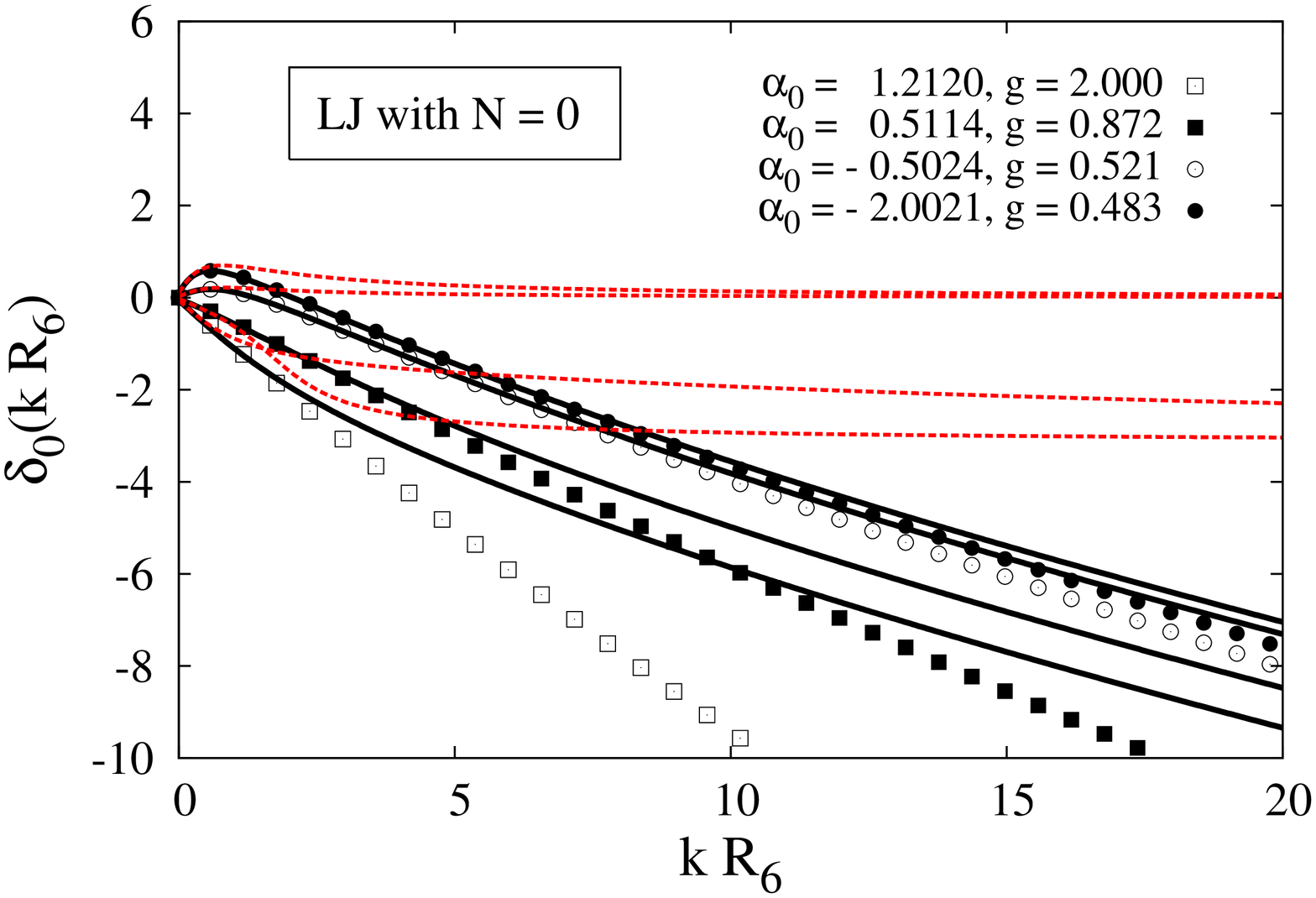,height=8cm,width=6cm,angle=270}
\epsfig{figure=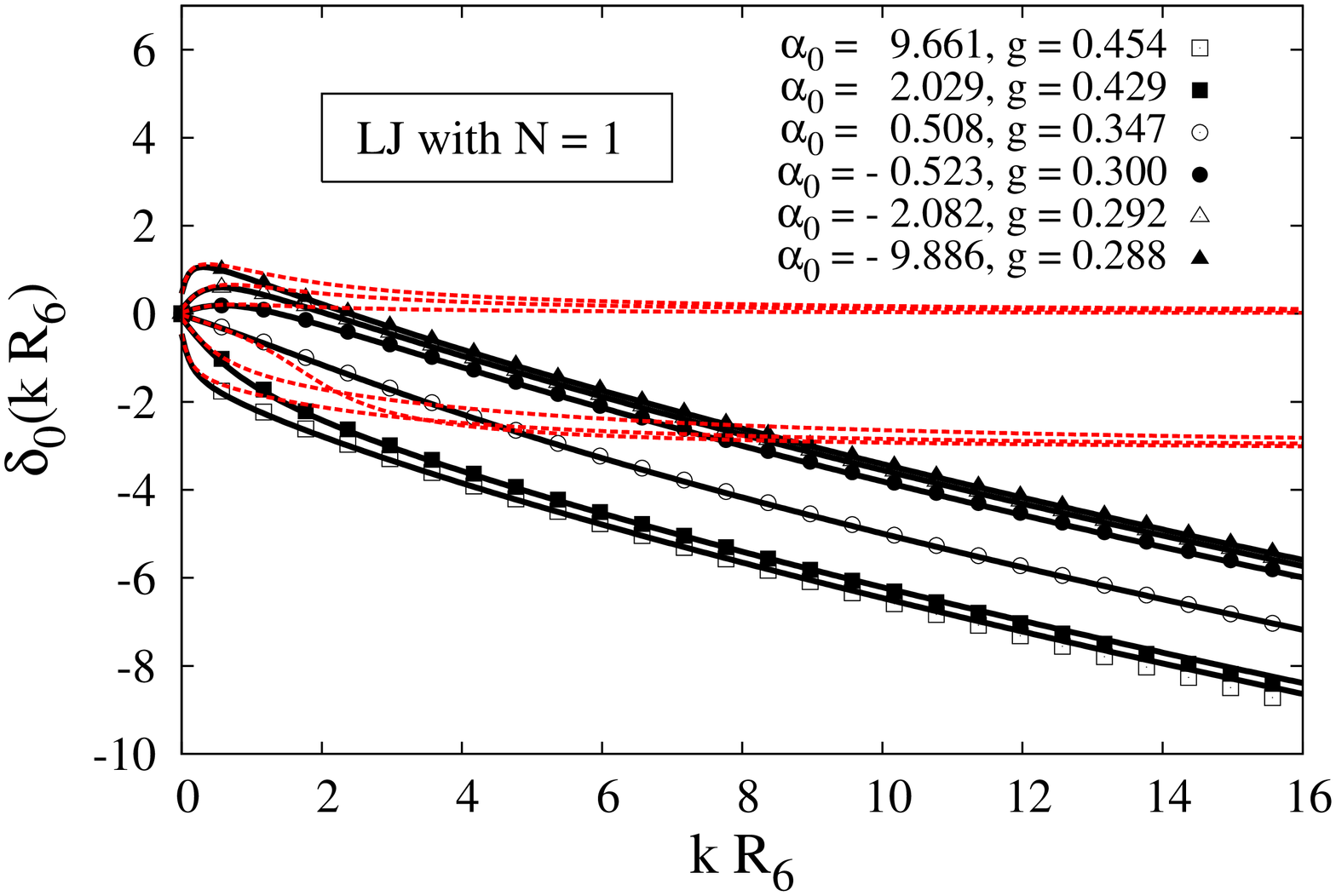,height=8cm,width=6cm,angle=270}
\epsfig{figure=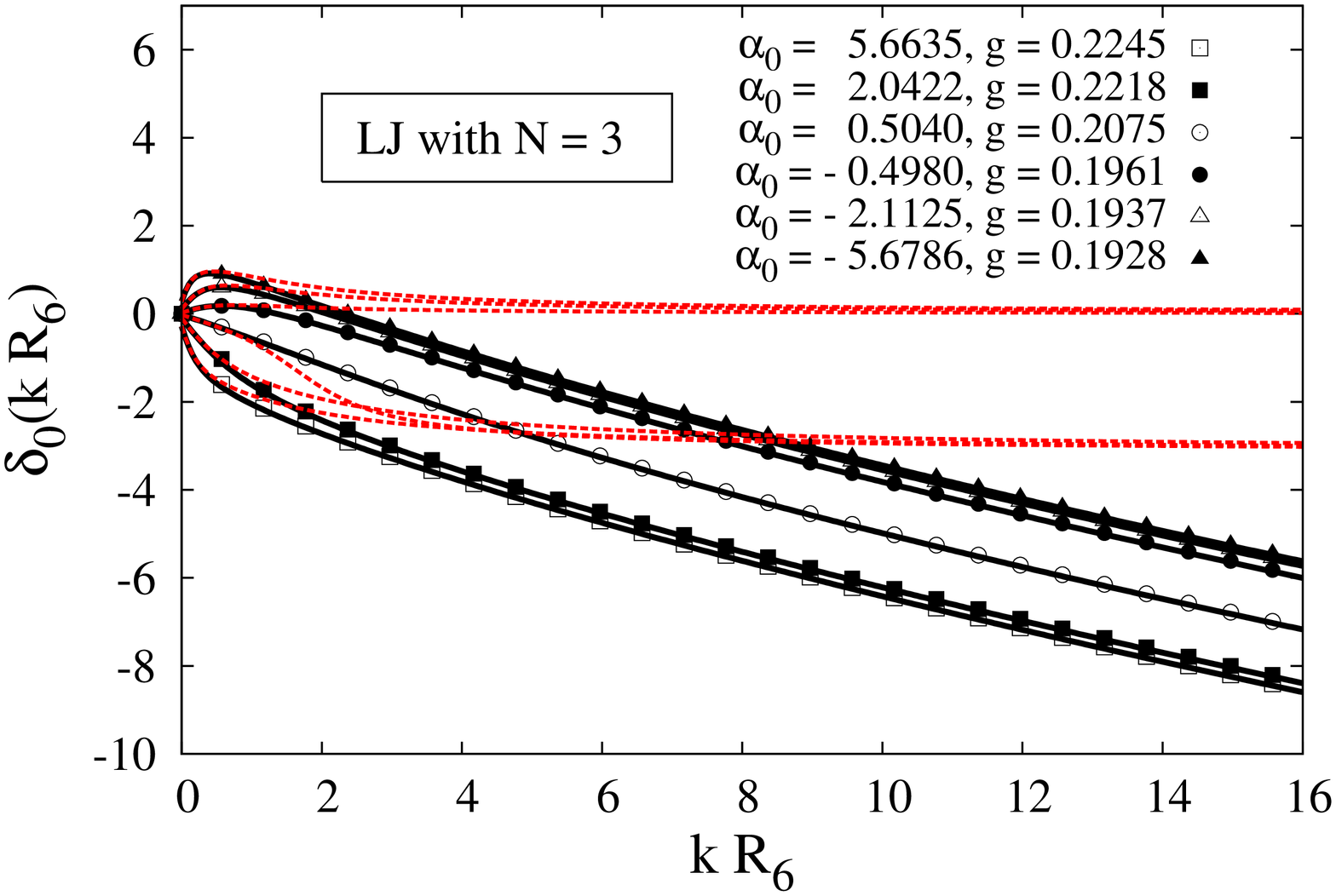,height=8cm,width=6cm,angle=270} 
\epsfig{figure=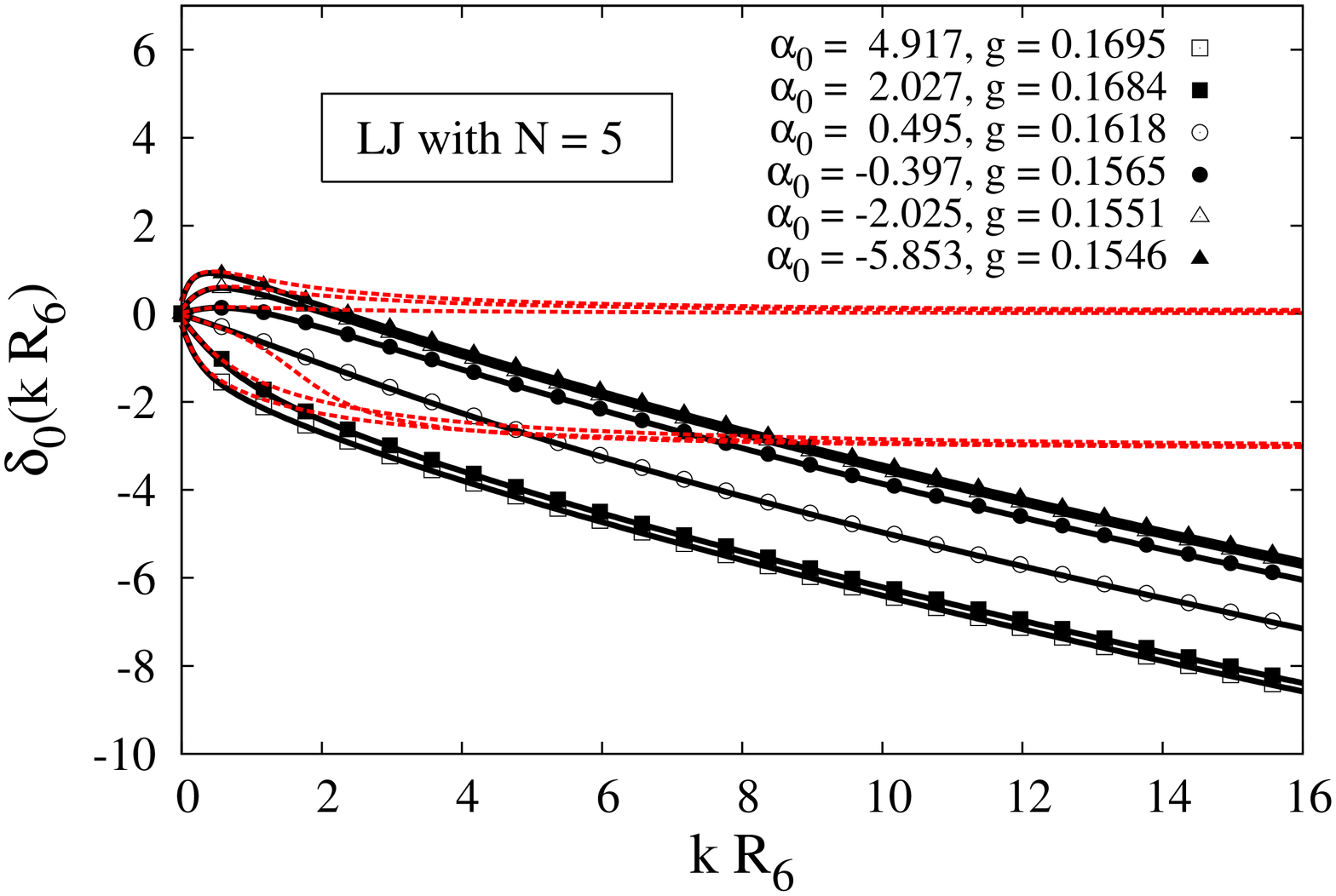,height=8cm,width=6cm,angle=270}
\end{center}
\caption{The Lennard-Jones s-wave phase shifts $\delta_0(k)$ (in radians) for
  different number of bound states $N=0,1,3,5$ as a function of the
  wave number (points) compared with the corresponding VdW renormalized ones
  having the same scattering length $\alpha_0$ (solid). We compare also with
  the effective range expansion truncated to second order $k \cot
  \delta_0 (k)= - 1/\alpha_0 + r_0 k^2/2 $ (dots).}
\label{fig:ps-LJ}
\end{figure*}

Using the VdW scaled units one can predict the scattering length
$\alpha_0$ from the LJ potential and the phase shifts unambiguously
for any value of $g$. The result is displayed in
Fig.~\ref{fig:LJ}. According to Levinson's theorem any time the
scattering length jumps from $-\infty$ to $+\infty$ a new bound state
dives from the continuum into the negative energy spectrum.  In
Fig.~\ref{fig:LJ} we display the effective range as given by
Eq.~(\ref{eq:r0_singlet}). The divergent values of $r_0$ correspond,
according to our low energy theorem, to points where the scattering
length goes through zero.  The very strong sensitivity to the precise
location of the position of the core is manifest. In Fig.~\ref{fig:LJ}
we compare the universal and renormalized VdW effective range formula
with the actual values deduced from the Lennard-Jones potential for a
different number of bound states $N=1,5,15$ (plus or minus one). The
minima in the curve corresponds to a value of the scattering length
passing from $-\infty$ to $+ \infty $ which corresponds to entering a
new bound state in the spectrum. As one sees in the LJ case there is a
multivalued function reflecting the multiple branches already observed
in Fig.~\ref{fig:LJ}. The rather universal pattern of this figure is
striking because it explicitly shows that to very small uncertainties
the value of the scattering length largely determines the value of the
effective range, {\it regardless} on the precise number of bound
states. The only remarkable exception corresponds to the case with no
bound states where the zero energy turning point is located at
increasingly large distances.

In a sense these correspond to almost low energy identical situations,
where additional bound states are hosted by the potential as the short
range repulsion is displaced towards the origin. Moreover, one expects
that the largest discrepancies from renormalized VdW and LJ should
take place in the case of large scattering lengths, since a larger
sensitivity to short distances is displayed in such a case. Actually,
this is what happens. We note that there is a relatively constant
shift between $r_0^{\rm LJ}$ and $r_0^{\rm VdW}$ of about $0.5-1.0$
for $|\alpha_0| > 10 R_6$. However, the renormalized theory works
better the larger the number of bound states and also for small
scattering lengths, since as we discussed in the previous section for
$\alpha_0 \gg R_6 $ large distances dominate.  A relevant question is
whether renormalization theory as explained above can account for the
behaviour of the phase shifts in a energy region where the De Broglie
wavelength is larger than the short distance $ k R_6 g \ll 1$
provided the scattering length $\alpha_0$ is the same. The result is
shown in Fig.~\ref{fig:ps-LJ} for a variety of values of $g$ which
cover several cases with large and small scattering lengths as
compared to the VdW radius. We see the anticipated similarity despite
the fact that both potentials are completely different at short
distances. Actually, the phase-shifts are indistinguishable for $k R_6
\ll 1 $, but they go hand in hand far beyond this expected region;
what matters is $ k R_6 g  \ll 1$. On the other hand, the ERE
given by Eq.~(\ref{eq:ere}) truncated to second order, i.e. taking
$v_2=0$ only works for $ k R_6 \ll 1$. In a sense this is equivalent
to ``seeing'' the Van der Waals force in a scattering experiment. The
point of renormalization theory is that it yields model independent
results and hence any discrepancy with data can be clearly attributed
to a deficient incorporation of the long distance physics.


\section{The Effective Field Theory and its limits}
\label{sec:photon-less}

At very low energies the interaction between neutral atoms can be
handled by an effective range expansion, Eq.~(\ref{eq:ere}), where the
long range character of the VdW potential becomes manifest in the
third term of the expansion. However, if only the first two terms are
retained
\begin{eqnarray}
k \cot \delta_0 (k)= - \frac1{\alpha_{0}}+ \frac12 r_{0} k^2 \, , 
\label{eq:ERE-trunc}
\end{eqnarray}
there arises the interesting possibility of universally representing
long range forces on equal footing with short range interactions.  We
may judge the quality of such an appealing approximation by comparing
Fig.~\ref{fig:ps-LJ} the result of the renormalized VdW theory with
the ERE. As we see the expansion truncated to second order breaks down
at low energies, namely $k R_6 \ll 1$, as expected.

Under these very restrictive conditions the problem can be
advantageously treated by using EFT methods, which are based on the
compelling idea that at such long wavelengths atoms behave as
elementary structureless particles. This point of view has been
stressed recently~(see e.g. \cite{Braaten:2004rn,Braaten:2007nq}) with
particular fruitful predictions in the three-body
problem~\cite{Braaten:2006vd,Platter:2009gz} where Efimov states have
been predicted.  This is usually done by considering the Lagrangian
density 
\begin{eqnarray}
&&{\cal L} (x) = \psi^\dagger (x) \left[ i \partial_t  +\frac{\nabla^2}{2m}
\right] \psi (x) \nonumber \\ &-&
 \frac12\int d^3 x' \psi^\dagger (x) \psi^\dagger (x') 
V(\vec x-\vec x') \psi(x') \psi(x)  \, , 
\label{eq:lagrangean}
\end{eqnarray}
where $\psi(x) \equiv \psi(\vec x,t)$ are space-time canonically
quantized fields with Fermi or Boson statistics depending upon the
spin nature of the atom as a whole.  The irreducible two-point
function is the potential $V(\vec x)$ which is taken as
\begin{eqnarray}
\langle \vec k' | V | \vec k \rangle \equiv \int d^3 x e^{i (\vec
  k-\vec k') \cdot \vec x } V( \vec x) = C_0 + C_2 (\vec k'^2 + \vec k^2)
\end{eqnarray}
Usually the coefficients $C_0$ and $C_2$ are assumed to be completely
arbitrary.  The corresponding Lippmann-Schwinger equation can be
solved requiring introducing a momentum space cut-off $\Lambda$ (see
e.g.  Ref.~\cite{Entem:2007jg} and references therein)
\begin{eqnarray} 
-\frac1{\alpha_0} &=& \frac{10(C_2 M \Lambda^3 -3)^2}{9 M \pi (
-C_2^2 M \Lambda^5 + 5 C_0)} - \frac{2\Lambda}{\pi} \, ,\\ \frac12
r_0 &=& \frac{50 C_2 \left( 3 + C_2 M \Lambda^3 \right)^2 \left( 6 + C_2 M 
\Lambda^3\right)}{27 \pi \left( -5 C_0 + C_2^2 \Lambda^5 M \right)^2  }  + \frac{2 }{\pi \Lambda} \, . 
\end{eqnarray} 
This leads for any cut-off $\Lambda$ to the mapping $(\alpha_0,r_0)
\to (C_0, C_2) $. Eliminating $C_0$ and $C_2$ in favour of $\alpha_0$ and
$r_0$, the phase shift becomes
\begin{eqnarray}
p \cot \delta (p) &=& - \frac{2 \Lambda}{\pi \alpha_0} \frac{(\pi - 2
  \Lambda \alpha_0)^2}{ 2 \Lambda (\pi - 2 \Lambda \alpha_0) +
  \alpha_0 p^2 (r_0 \pi \Lambda -4)} \nonumber \\ &-&
\frac{2\Lambda}{\pi} + \frac{p}{\pi}\log \frac{\Lambda+p}{\Lambda-p}
\, ,
\label{eq:ps-Lambda}
\end{eqnarray}
which is a {\it real} quantity for $p < \Lambda$, meaning that two-body unitarity is
fulfilled.  However, one has complex solutions for $C_0$ and $C_2$ if
\begin{eqnarray}
\alpha_0^2 r_0 \pi \Lambda^3 - 16 \alpha_0^2 \Lambda^2 + 12 \alpha_0
\pi \Lambda -3 \pi^2 \le 0 \, . 
\label{eq:caus-mom}
\end{eqnarray}
which means that the effective Lagrangian, Eq.~(\ref{eq:lagrangean}),
becomes non hermitian ${\cal L}^\dagger (x)\neq {\cal L}(x)$. On the
other hand, it is well known that three-body unitarity rests on
off-shell two-body unitarity~\cite{PhysRev.135.B1225}, a condition
which cannot be met if the interaction is not hermitian, mainly
because Schwartz's principle. So the lesson is that while a complex
two-body potential may fulfill on-shell two-body unitarity, a
violation of three body unitarity is still
possible~\cite{Entem:2007jg}.

\begin{figure}
\epsfig{figure=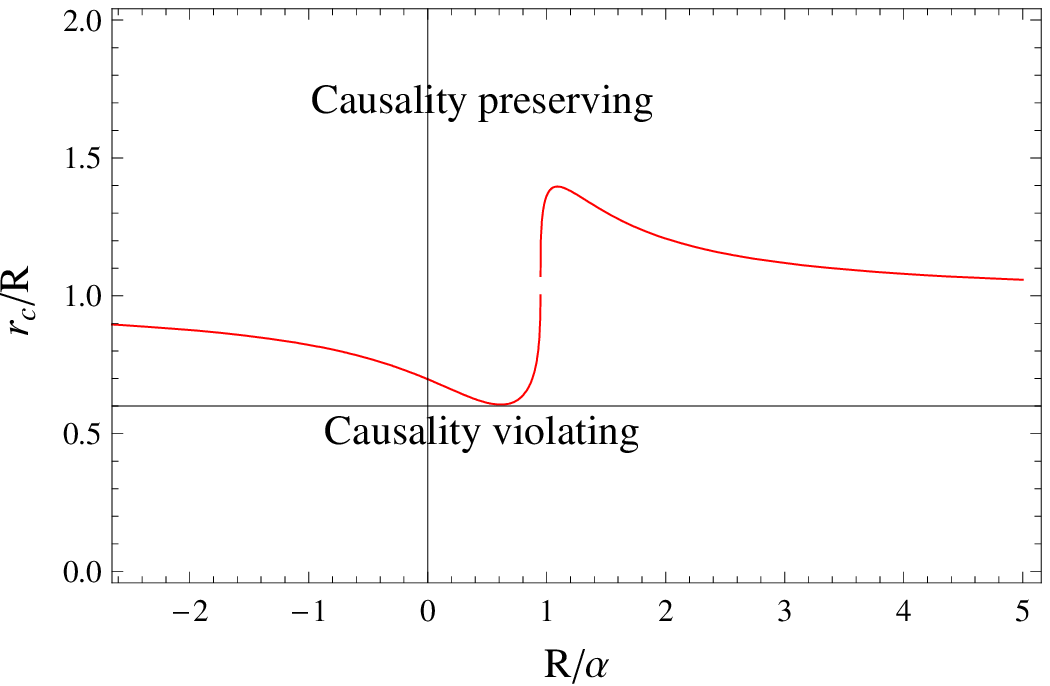,height=4.5cm,width=4cm,angle=0} 
\epsfig{figure=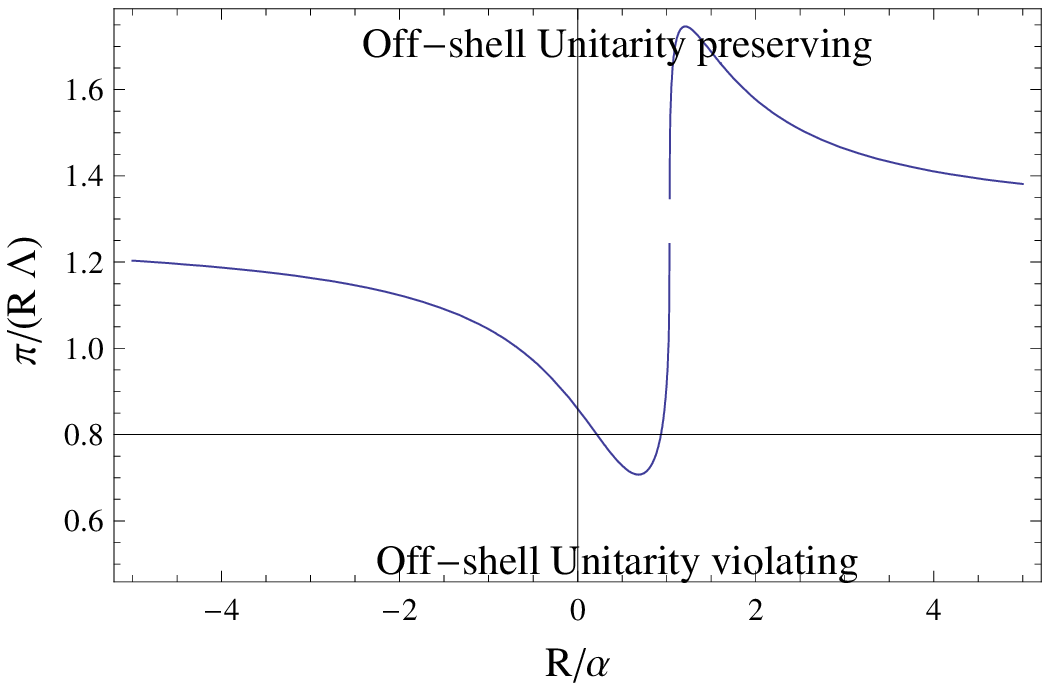,height=4.5cm,width=4cm,angle=0}
\caption{Boundaries on the cut-off for Van der Waals systems with
  reduced potential $2 \mu V(r) = - R^4 /r^6$ for which causality in
  coordinate space (left panel) and off-shell unitarity in momentum space
  (right panel) is preserved.}
\label{fig:causality}
\end{figure}

Generally, Eq.~(\ref{eq:caus-mom}) imposes a limit on the maximum
value of the cut-off $\Lambda$, but to find it we need to know both
$\alpha_0$ and $r_0$.  For our case of VdW interactions we have seen
that the universal formula for $r_0$ in terms of $\alpha_0$,
Eq.~(\ref{eq:r0-ggf}) works extremely well if we judge by
Fig.~\ref{fig:atoms-curve}. Thus, if we merge Eq.~(\ref{eq:caus-mom})
equal to null and Eq.~(\ref{eq:r0-ggf}) we obtain a boundary in the
$(\Lambda, \alpha_0)$ plane which is suitably represented in
Fig.~\ref{fig:causality}. Essentially and up to minor variations the
meaning is that the cut-off $\Lambda$ cannot exceed the VdW wave
number, $ \Lambda < \pi / ( 2 R_6) $. This issue is relevant for
three-body calculations~\cite{Platter:2006ev,Platter:2008cx} as
addressed recently~\cite{Canham:2009xg}.

The analysis of the problem in coordinate space {\it assuming} an
effective local and energy independent long distance dynamics and an
energy dependent boundary condition at short distances have been
analyzed in Ref.~\cite{PavonValderrama:2005wv} for s-waves and in
Ref.~\cite{Cordon:2009pj} in the three-dimensional case. It is found
that the locality condition for an s-wave implies
\begin{eqnarray}
\frac{d}{d k^2} \left[\frac{u_{k,{\rm S}}'(r_c)}{u_{k,{\rm
S}}(r_c)} \right] = - \frac{
\int_0^{r_c} u_{k,{\rm S}} (r)^2 dr}{u_{k,{\rm S}} (r_c)^2} \le 0
\end{eqnarray} 
where $u_{k,{\rm S}}(r)$ is the wave function for $r \le r_c$. Note
the resemblance with Eq.~(\ref{eq:positivity}). If there would be no
interaction for $r > r_c$ then we have
\begin{eqnarray}
u_{k} (r) = \sin \left( k r + \delta_0 (k) \right) \, ,
\quad r > r_c \, , 
\end{eqnarray} 
which can be matched to the inner $r < r_c$ region yielding 
\begin{eqnarray}
\frac{d}{d k^2} \left[ k \cot \left( k r_c + \delta_0(k) \right)
  \right] \le 0 \, ,
\end{eqnarray}
this is equivalent to Wigner's causality
condition~\cite{Wigner:1955zz}, as noted in \cite{Phillips:1996ae},
and combined to the ERE, Eq.~(\ref{eq:ERE-trunc}), provides a
constraint on the effective range
\begin{eqnarray}
r_0 / (2 r_c)  \le 1 - r_c/\alpha_{0}+ r_c^2/ (3 \alpha_0^2)  \, .  
\end{eqnarray}
If there is no potential for $r> r_c$ and take $r_{0}= r_{\rm VdW}$
we get a universal lower limit for $r_c$.  A similar conclusion has been
presented recently~\cite{Hammer:2009zh}.  The conditions featuring
causality in coordinate space as well as off-shell unitarity in momentum
space are depicted in Fig.~\ref{fig:causality}, and suggest that
modelling a finite range of VdW forces by an effective Lagrangian such
as Eq.~(\ref{eq:lagrangean}) requires assuming a cut-off distance
larger than the VdW length.

\section{Conclusions}

Renormalization ideas can profitably be exploited in conjunction with
the superposition principle of boundary conditions in the description
of model independent and universal features of the VdW force. Our main
points are

\begin{itemize}

\item Van der Waals interactions between neutral atoms obey scaling
  rules which allow to determine the scattering and binding properties
  universally. They are well satisfied phenomenologically and extend
  much beyond low energy approximations such as the effective range
  expansion.

\item There is a clear dominance of the leading $C_6$ contribution in
  a rather wide energy range. The range where higher order corrections
  due to $C_8$ or $C_{10}$ provide a distinct correction yet the
  finite size effects can still be neglected is extremely narrow or
  inexistent.

\item Van der Waals potentials can be represented by short distance
  contact interactions under restrictive conditions based on causality
  and off-shell unitarity which are independent on the value of the
  scattering length. It is inconsistent to model VdW forces assuming a
  cut-off distance smaller than the VdW length.

\end{itemize}

We thank M. Pav\'on Valderrama and R. Gonz\'alez F\'erez for
discussions.  Work supported by Spanish DGI and FEDER funds
with grant FIS2008-01143, Junta de Andaluc{\'\i}a grant FQM-225-05, EU
Integrated Infrastructure Initiative Hadron Physics Project contract
RII3-CT-2004-506078.
%


\end{document}